# THE BEHAVIOR OF THE OPTICAL AND X-RAY EMISSION FROM SCORPIUS X-1

B. J. McNamara,[1] T. E. Harrison,[1] R. T. Zavala, Eduardo Galvan, Javier Galvan, T. Jarvis,
GeeAnn Killgore, O. R. Mireles, D. Olivares, B. A. Rodriquez, M. Sanchez,
Allison L. Silva, Andrea L. Silva, and E. Silva-Velarde
Department of Astronomy, New Mexico State University, 1320 Frenger Mall, Las Cruces, NM 88003-8001;
bmcnamar@nmsu.edu, tharriso@nmsu.edu, rzavala@nmsu.edu

AND

M. R. Templeton
American Association of Variable Star Observers, 25 Birch Street, Cambridge, MA 02138-1205

## ABSTRACT

In 1970, Hiltner & Mook reported the results of the first multiyear study of the optical emission from Sco X-1. They found that the Sco X-1 $B$-magnitude histograms changed from year to year. Subsequent multiwavelength campaigns confirmed the variable nature of these optical histograms and also found that the X-ray and optical emissions were only correlated when Sco X-1 was brighter than about $B = 12.6$. Models had suggested that the optical emission from this source arose from X-rays reprocessed in an accretion disk surrounding the central neutron star. It was therefore difficult to explain why the optical and X-ray fluxes were not more closely correlated. In 1994 and 1995, two new simultaneous optical and X-ray campaigns on Sco X-1 were conducted with the Burst and Transient Source Experiment on the *Compton Gamma Ray Observatory* and the 1 m Yale telescope at Cerro Tololo Inter-American Observatory. Using these data and models by Psaltis, Lamb, & Miller, it is now possible to provide a qualitative picture of how the X-ray and optical emissions from Sco X-1 are related. Differences in the $B$-magnitude histograms are caused by variations in the mass accretion rate and the relatively short time period typically covered by optical investigations. The tilted-$\Gamma$ pattern seen in plots of the simultaneous X-ray and optical emission from Sco X-1 arises from (1) the nearly linear relation between the optical $B$ magnitude and the mass accretion rate in the range $13.3 \geq B \geq 12.3$ and an asymptotic behavior in the $B$ magnitude outside this range, and (2) a double-valued relation between the X-ray emission and mass accretion rate along the normal branch and lower flaring branch of this source.

## 1. INTRODUCTION

Scorpius X-1 is a member of a group of objects called Z sources. This name arises from the fact that when the soft X-ray colors of one of these objects are plotted against its hard X-ray colors, a Z-shaped pattern is produced. Because of this distinctive feature, this type of plot is frequently referred to as a Z diagram. The top portion of the Z is called the horizontal branch (HB), the middle segment is referred to as the normal branch (NB), and the bottom part is called the flaring branch (FB). The location along this Z pattern is normally given by a quantity $S_Z$. Different conventions exist for measuring $S_Z$, but its value always increases from the HB to the FB (Hertz et al. 1992; Dieters & van der Klis 2000). Multiwavelength studies of Z sources, such as Sco X-1 and Cyg X-2, show that the mass accretion rate, ultraviolet line and continuum fluxes, and optical brightness are correlated with the value of $S_Z$ (Hasinger et al. 1990; Vrtilek et al. 1990, 1991).

The optical emission from Sco X-1 arises from reprocessed X-rays created when mass is transferred from the system's low-mass secondary star to its neutron star. Using *International Ultraviolet Explorer* data, Vrtilek et al. (1990, 1991) investigated whether this reprocessing site was associated with the surface of the secondary star or the accretion disk. They found that the dominant site is the accretion disk and estimated that the accretion rate monotonically increased along the HB to the FB from $0.4 \times 10^{-8}$ to $1.2 \times 10^{-8}$ $M_\odot$ yr$^{-1}$. Data from their paper suggest that the accretion rate and UV continuum flux are related as $dM/dt = 0.68$ UV(1224–1986 Å) $- 1.26$, where $M$ is measured in units of $10^{-9}$ $M_\odot$ yr$^{-1}$ and the UV flux is measured in units of $10^{-12}$ ergs cm$^{-2}$ s$^{-1}$. Willis et al. (1980) have shown that changes in the UV flux are tracked by changes in the Johnson $B$ magnitude. A linear relation must therefore also exist between $dM/dt$ and $B$. A similar relation between the accretion rate and the X-ray emission does not exist, because the optical and X-ray emissions do not always vary in the same manner. When Sco X-1 is optically bright, the X-ray and optical fluxes are directly correlated; at fainter optical magnitudes they can be either correlated or uncorrelated, while at still fainter optical magnitudes, these fluxes again appear to be directly correlated. This complex relation causes simultaneously obtained Sco X-1 X-ray and optical magnitude histograms to have dramatically different morphologies. In an optical and X-ray study conducted in 1970 using *Vela* spacecraft data, Mook et al. (1975) found that Sco X-1 possessed a trimodal histogram of $B$ magnitudes, whereas only a single strong peak was present in its X-ray histogram. One year later, Bradt et al. (1975) collected data from *Uhuru* and a network of optical observatories and found that Sco X-1 then possessed a bimodal histogram of $B$ magnitudes and an X-ray histogram that possessed only a

---

single strong peak with a sparsely populated tail at higher count rates. Using *OSO 7* satellite data and optical data obtained in 1972, Canizares et al. (1975) found similar results to those of Bradt et al., but the peaks in their Sco X-1 optical histogram were at different *B* magnitudes. In a fifth multiwavelength study, Ilovaisky et al. (1980) used *Copernicus* and *SAS 3* data obtained from 1975 to 1977 and found that Sco X-1 then possessed a trimodal distribution of *B* magnitudes and a single-peaked X-ray histogram.

Psaltis, Lamb, & Miller (1995) have developed a model that allows the emission from a rapidly rotating magnetic neutron star that is accreting mass from a companion star at 0.5–1.1 times the Eddington critical rate to be computed. Three emission sites are considered in their model: the neutron star magnetosphere, a hot central corona (HCC), and an extended corona through which material flows radially toward the neutron star. Radiation produced in the magnetosphere is approximated as a blackbody with a high-energy cutoff. As this radiation enters the HCC, the spectrum is altered as a result of changes in the electron scattering opacity ($\tau_{es}^{HCC}$). This spectrum is further modified in the extended corona by changes in that region's electron temperature and electron scattering opacity. To simply their calculations, Psaltis et al. assume that the electron temperatures in the magnetosphere and HCC are equal and that, in agreement with observations, $\tau_{es}^{HCC} = 6$. The temperature in the radial flow is then determined by solving the electron kinetic equation. To compare their model results with observed count rate spectra, Psaltis et al. also assume that a source's luminosity and accretion rate are directly related and that the source's distance can be obtained by equating the flux level at the intersection of its FB and NB to the Eddington luminosity. These simplifying assumptions then allow a source's X-ray count rate spectrum, Z plot, and hard color versus count rate relation to be compared with observational data.

Because of the complex nature of the emission from Sco X-1, we decided to conduct another multiwavelength campaign on this source. The specific questions we wished to investigate were the following: (1) How are the X-ray emission and *B* magnitudes of Sco X-1 related? (2) Can the Sco X-1 *B*-magnitude histogram be understood in terms of other properties of this system, and why is it so variable? (3) How do these findings relate to current models of low-mass X-ray binaries?

## 2. OBSERVATIONS AND DATA REDUCTION

The new X-ray data employed in this study consist of *Compton Gamma Ray Observatory* (*CGRO*) BATSE Spectroscopy Detector (SD) step fluxes. Operational from 1991 to 2000, the BATSE SD system consisted of eight detectors arranged on *CGRO* so that they provided an all-sky monitoring capability. Each SD consisted of a 7.6 cm thick NaI(TI) scintillator encased in a shielded frame with a 146 cm$^2$ beryllium entrance window. Measurements were collected over a wide range of energy (approximately 8 keV to 11 MeV, depending on the gain setting), but this study only uses the SD 8–16 keV measurements. Each measurement lasted 2.048 s. An extensive discussion of this instrument can be found in Schaefer et al. (1992), McNamara, Harmon, & Harrison (1995), and McNamara et al. (1998).

As *CGRO* orbited Earth, an SD would alternately see Sco X-1 rise and set below Earth's limb. At these times the observed X-ray flux would suddenly increase and decrease, creating what is referred to as a "step flux" in this paper. A detailed discussion of how these step fluxes are measured and the corrections needed to account for the differing *CGRO* pointing angles and detector sensitivities can be found in McNamara et al. (1998). The average error of a step flux in that study was $3.7 \pm 0.8$ counts. The signal-to-noise ratio of the step fluxes used in this study varies from approximately 6.0 (when Sco X-1 is near its X-ray minimum) to about 18.0 (when Sco X-1 is near the bright end of its FB). Under optimal circumstances, the time between consecutive Sco X-1 rise step fluxes was 5600 s and the rise-set interval was about 3300 s. From a single ground-based site, about a dozen opportunities existed to obtain coincident step fluxes and optical data during the course of a night.

The optical data used in this study were obtained with the 1 m Yale telescope and the People's Photometer at CTIO. In 1994, 15 nights of photoelectric Johnson *B* and *V* data were obtained with integration times of 2 and 4 s. In 1995, 11 nights of additional data were collected, but during this period only the Johnson *B* filter was used and all integration times had a duration of 1 s. The *B* and *V* measurements were then transformed into differential magnitudes with respect to two nearby reference stars. The coordinates and magnitudes of these stars are listed in Table 1. The error for a 1 s Sco X-1 *V* observation was 0.03 mag. The error for a 1 s measurement in *B* was 0.02 mag. Table 2 provides a log of the optical data. The third column provides the nightly optical extinction coefficients (in magnitudes per air mass), followed by the number of hours Sco X-1 was monitored, the integration time (in seconds), and the number of Sco X-1 observations. Optical observations were continuously acquired while Sco X-1 was visible from *CGRO*. However, this study only utilizes simultaneous data collected during a Sco X-1 SD rise or set.

## 3. *B*-MAGNITUDE HISTOGRAMS

Figure 1 (*left*) shows histograms of the *V* and *B* magnitudes obtained during our 1994–1995 campaigns. In 1994,

TABLE 1
Reference-Star Information

| Object | $\alpha$ (J2000) | $\delta$ (J2000) | $V$ | $B-V$ |
| --- | --- | --- | --- | --- |
| Sco X-1 | 16 19 55.07 | −15 38 24.8 | ∼12.1–13.4 | ∼0.0–0.2 |
| Primary reference | 16 20 10.60 | −15 39 43.3 | 11.30 | 0.66 |
| Secondary reference | 16 19 40.90 | −15 37 25.3 | 11.47 | 1.26 |

Note.—Units of right ascension are hours, minutes, and seconds, and units of declination are degrees, arcminutes, and arcseconds.

TABLE 2
CTIO Observational Log

| UT | Filter | Extinction | Hours | $T$ (s) | No. Obs. |
|---|---|---|---|---|---|
| 1994 May 12............ | $B$ | 0.18 | 7.3 | 4 | 977 |
|  | $V$ | 0.14 | 7.2 | 4 | 3830 |
| 1994 May 13............ | $B$ | 0.23 | 8.2 | 4 | 203 |
|  | $V$ | 0.13 | 8.0 | 4 | 4370 |
| 1994 May 14............ | $B$ | 0.26 | 7.9 | 4 | 437 |
|  | $V$ | 0.14 | 7.5 | 4 | 3928 |
| 1994 May 15............ | $B$ | 0.25 | 8.8 | 4 | 380 |
|  | $V$ | 0.13 | 8.4 | 4 | 5658 |
| 1994 May 16............ | $B$ | 0.25 | 7.7 | 4 | 243 |
|  | $V$ | 0.13 | 8.3 | 4 | 5521 |
| 1994 May 19............ | $B$ | 0.20 | 6.9 | 4 | 149 |
|  | $V$ | 0.12 | 6.9 | 4 | 4799 |
| 1994 May 20............ | $B$ | 0.24 | 8.6 | 4 | 148 |
|  | $V$ | 0.14 | 8.5 | 4 | 4846 |
| 1994 May 21............ | $B$ | 0.27 | 7.5 | 4 | 199 |
|  | $V$ | 0.15 | 7.5 | 4 | 4770 |
| 1994 May 22............ | $B$ | 0.28 | 4.2 | 4 | 117 |
|  | $V$ | 0.16 | 4.2 | 4 | 2520 |
| 1994 Jun 3 .............. | $B$ | 0.29 | 9.4 | 2 | 10986 |
|  | $V$ | 0.18 | 8.8 | 2 | 523 |
| 1994 Jun 5 .............. | $B$ | 0.28 | 5.8 | 2 | 5982 |
|  | $V$ | 0.18 | 5.5 | 2 | 242 |
| 1994 Jun 6 .............. | $B$ | 0.29 | 5.9 | 2 | 5078 |
|  | $V$ | 0.20 | 5.9 | 2 | 118 |
| 1994 Jun 8/9 .......... | $B$ | 0.24 | 9.1 | 2 | 8039 |
|  | $V$ | 0.13 | 8.7 | 2 | 97 |
| 1994 Jun 9/10......... | $B$ | 0.28 | 9.1 | 2 | 8860 |
|  | $V$ | 0.16 | 9.1 | 2 | 190 |
| 1994 Jun 10/11....... | $B$ | 0.30 | 8.7 | 2 | 7369 |
|  | $V$ | 0.21 | 8.5 | 2 | 130 |
| 1994 Jun 11/12....... | $B$ | 0.18 | 8.8 | 2 | 7433 |
|  | $V$ | 0.28 | 8.6 | 2 | 121 |
| 1994 Jun 12/13....... | $B$ | 0.06 | 9.2 | 2 | 6521 |
|  | $V$ | 0.15 | 9.1 | 2 | 116 |
| 1995 May 2 ............ | $B$ | 0.25 | 6.5 | 1 | 7025 |
| 1995 May 3 ............ | $B$ | 0.20 | 7.2 | 1 | 10102 |
| 1995 May 4 ............ | $B$ | 0.20 | 7.1 | 1 | 11296 |
| 1995 May 8 ............ | $B$ | 0.19 | 8.1 | 1 | 9615 |
| 1995 May 9 ............ | $B$ | 0.25 | 7.9 | 1 | 9323 |
| 1995 May 20........... | $B$ | 0.28 | 8.5 | 1 | 6232 |
| 1995 May 21........... | $B$ | 0.25 | 8.5 | 1 | 6369 |
| 1995 May 22........... | $B$ | 0.25 | 7.8 | 1 | 7566 |
| 1995 May 23........... | $B$ | 0.25 | 8.2 | 1 | 7642 |
| 1995 May 24........... | $B$ | 0.25 | 7.8 | 1 | 7374 |
| 1995 May 26........... | $B$ | 0.30 | 7.2 | 1 | 5700 |

Sco X-1 was observed in both these filters, whereas in 1995 it was observed only in $B$. During these two observing seasons, Sco X-1 possessed three optical states: a bright, an intermediate, and a faint state. Gaussian fits to these histograms using the IRAF routine SPLOT are shown in the right panels of Figure 1. Table 3 lists the peak $B$ magnitudes based on these fits. In 1994, the percentage of $B$ observations in each of these three states was 36%, 22%, and 42%, respectively. In 1995 these numbers were 47%, 13%, and 40%, although the faint state was about 0.3 mag brighter than in 1994. During the 1994 campaign, the bright state had a color of $B-V \sim 0.04$, whereas the intermediate and faint states had $B-V$ of $\sim 0.05$ and $\sim 0.17$, respectively. The fact that Sco X-1 does not appreciably change in $B-V$ when moving from its bright to its faint state has been noted by other investigators (see Mook 1967).

The 1994 and 1995 peak $B$ magnitudes can be compared with values found in the prior studies of Hiltner & Mook (1970), Canizares et al. (1975), Mook et al. (1975), Bradt et al. (1975), and Ilovaisky et al. (1977). This comparison is provided in Figure 2 and Table 3. Peak $B$ magnitudes in the 1967, 1968, 1969, 1970, 1972, and 1977 histograms were identified in the same manner as for the 1994–1995 data. Peaks in the 1971 histogram are based on visual estimates because $B$ magnitudes in Bradt et al. (1975) were presented in a logarithmic plot that made reconstruction of a histogram like those shown in Figure 1 impossible. Five statements can be made based on all of this information: (1) during an observing season Sco X-1 can occupy discrete brightness states, but the peak $B$ magnitudes of these states are generally not the same from year to year; (2) the amount of time Sco X-1 spends in its bright, intermediate, and faint states varies from season to season; (3) the peak $B$ brightness of Sco X-1 appears to have increased by approximately 0.2 mag since the early 1970s; (4) Sco X-1 varies between $B = 12.1$ and $B = 13.4$ (a factor of 3.3 in brightness), and its $B-V$ color increases from 0.04 to 0.17 as its brightness decreases; and (5) the most stable Sco X-1 $B$-magnitude peak appears to be located at $B \sim 12.60$. This state is present in eight of the nine entries in Table 3. None of the other $B$-magnitude peaks are likewise present over long periods of time. Based on the simultaneous X-ray data available in the Bradt et al., Canizares et al., and Ilovaisky et al. studies, the $B \sim 12.5$ peaks probably correspond to the $B \sim 12.3$ peak found in the 1994–1995 campaigns.

## 4. CORRELATED X-RAY AND OPTICAL MAGNITUDES

Figure 3 shows the simultaneous BATSE SD step fluxes and $B$ magnitudes obtained during our 1994 and 1995 campaigns. The tilted-Γ pattern evident in this diagram is similar to that found in the studies of Canizares et al. (1975; their Fig. 5), Bradt et al. (1975; their Fig. 9), and Mook et al. (1975; their Fig. 3). The X-ray energy ranges examined in those three studies were 3–10, 2.4–6.9, and 3–12 keV, respectively, and are lower than the new 8–16 keV BATSE SD measurements presented here. A similar diagram was provided by Ilovaisky et al. (1980), but it was based on only 3 days of data and is therefore not comparable. An examination of these $B$ versus X-ray magnitude plots suggests that Sco X-1 occupies three states. When Sco X-1 is brighter than $B \sim 12.5$, changes in the X-ray and optical magnitudes are directly related, and the range in the X-ray magnitudes is large compared with the optical variation. In the magnitude interval $13.4 \geq B \geq 12.7$, the $B$-magnitude range is larger than the range in X-ray magnitudes, and changes in these magnitudes are inversely related. Finally, below $B \sim 13.4$ the optical and X-ray magnitudes are positively related, and the range of X-ray magnitudes is larger than the range of $B$ magnitudes. It is difficult to precisely determine these relations, because the data are fairly limited. Table 4 contains values of $\Delta B/\Delta X$ estimated from a visual inspection of Figure 3 and the similar figures found in the above papers. Since the calculation of these ratios involves subjective judgements about the location of the state boundaries, these slopes should only be considered as suggestive. Figure 3 from Mook et al. (1975) illustrates the need for a greater number of longer term simultaneous optical and X-ray measurements. During their campaign, Sco X-1 spent most of its time near three distinct *B* magnitudes, and the transition between these states was poorly sampled. Clearly, longer term measurements, which will allow these transitions to be better mapped, are required to define how Sco X-1 moves in this diagram. Therefore, the entries in Table 4 are only intended to demonstrate that when Sco X-1 is optically bright, large and correlated changes in its X-ray flux are to be expected, whereas the opposite situation occurs when Sco X-1 is in its intermediate optical brightness state.

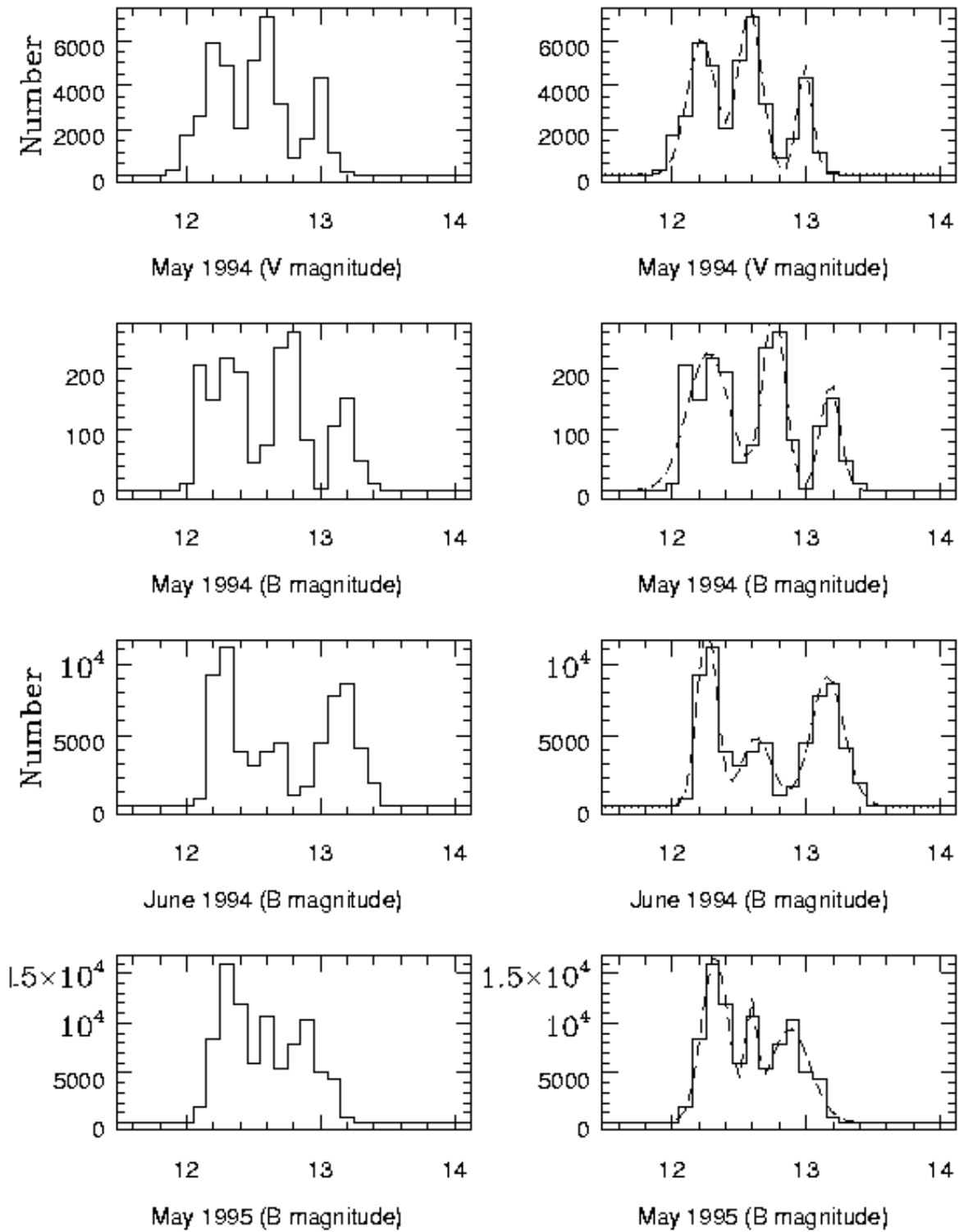

Fig. 1.—The *B*- and *V*- magnitude histograms based on data collected during our 1994 and 1995 campaigns using the 1 m Yale telescope at CTIO. Dashed lines in the right panels are Gaussian fits to the histograms using the IRAF routine SPLOT. Note that three states appear to be present: a faint state, an intermediate-brightness state, and a bright state. The locations of these states should be contrasted with those seen in Fig. 2.

TABLE 3
B-MAGNITUDE HISTOGRAM PEAKS

| DATE | \multicolumn{5}{c}{PEAK B MAGNITUDES} | DATA SOURCE |
| | 1 | 2 | 3 | 4 | 5 | |
|---|---|---|---|---|---|---|
| 1967 (21 days): | | | | | | |
| Apr 11–14, 17–20, 23, 30, May 1–6, 16–20 | 12.45 | 12.64 | 12.85 | 13.01 | 13.23 | Hiltner & Mook 1967 |
| 1968 (8 days): | | | | | | |
| Apr 2, May 8–9, 15, 17–20 | 12.50 | ... | 12.89 | 13.08 | 13.38 | Hiltner & Mook 1970 |
| 1969 (12 days): | | | | | | |
| May 7–16, 23, 25 | ... | 12.64 | ... | ... | ... | Hiltner & Mook 1970 |
| 1970 (23 days): | | | | | | |
| Apr 26–May 18 | ... | 12.60 | 12.92 | ... | 13.19, 13.50 | Mook et al. 1975 |
| 1971 (19 days): | | | | | | |
| Feb 25–28, Mar 1–3, 22–31, Apr 1–2 | 12.45 | 12.6 | ... | 13.1 | 13.4 | Bradt et al. 1975 |
| 1972 (9 days): | | | | | | |
| Jun 6–14 | 12.55 | 12.7 | ... | ... | 13.3 | Canizares et al. 1975 |
| 1977 (3 days): | | | | | | |
| Mar 14–16 | 12.48 | 12.7 | ... | 13.05 | ... | Ilovaisky et al. 1980 |
| 1994 (15 days): | | | | | | |
| May 12–16, 19–22, Jun 4–5, 8–11 | ... | 12.63 | ... | 13.16 | ... | This study |
| 1995 (12 days): | | | | | | |
| May 1–4, 8–9, 19–23, 25 | 12.59 | 12.88 | ... | ... | ... | This study |

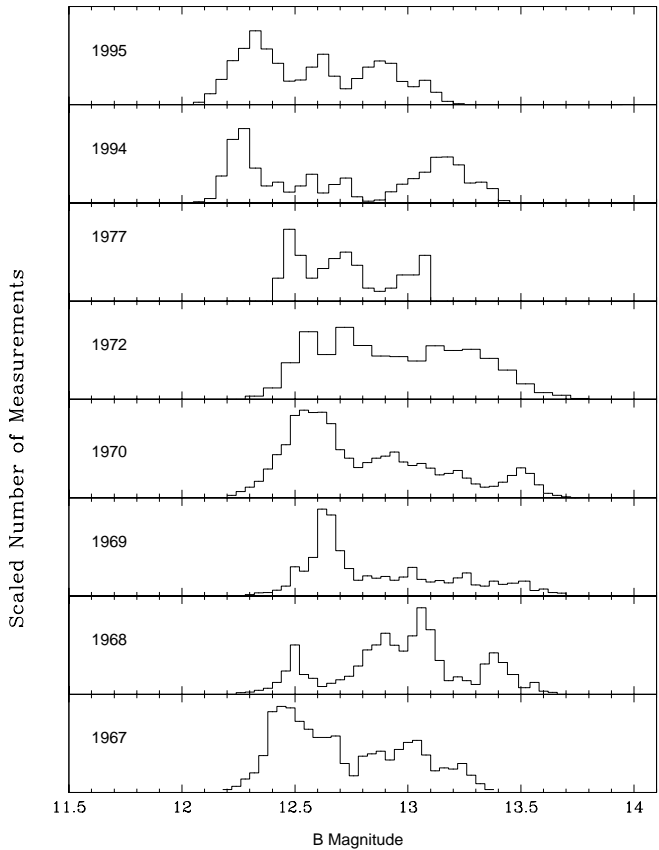

FIG. 2.—A survey of Sco X-1 B-magnitude histograms. *Top to bottom*, our new 1995 and 1994 CTIO B-band data, 1977 data from Ilovaisky et al. (1980), 1972 data from Canizares et al. (1975), 1970 data from Mook et al. (1975), 1969 and 1968 data from Hiltner & Mook (1970), and 1967 data from Hiltner & Mook (1967). The peak B magnitudes clearly change from year to year.

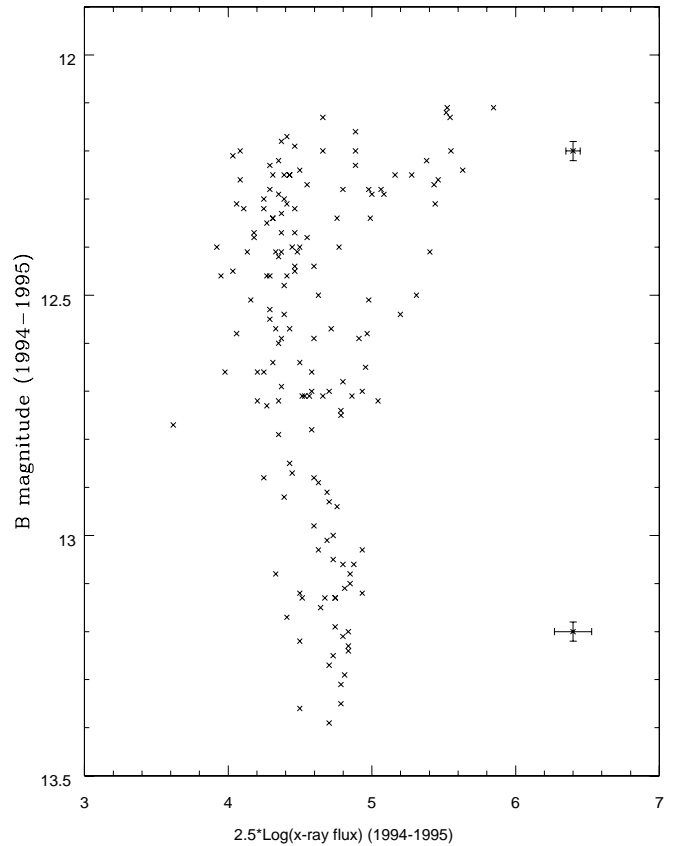

FIG. 3.—Simultaneous optical B and BATSE SD X-ray magnitudes. When Sco X-1 has $B < 12.5$, changes in its X-ray magnitude are larger than changes in the B magnitude, and these changes are positively correlated. When Sco X-1 has $B > 12.5$, changes in its B magnitude are larger than changes in its X-ray magnitude, and these changes are anticorrelated. Typical error bars for the B magnitudes and BATSE SD magnitudes are shown at lower right.

TABLE 4
$\Delta B/\Delta X$ for Scorpius X-1's Three Brightness States

| Source | Bright | Intermediate | Faint |
|---|---|---|---|
| This study | 0.05 | −2.5 | a |
| Mook et al. 1975 | 0.06 | −6.1 | 0.12 |
| Canizares et al. 1975 | 0.34 | −2.9 | a |
| Bradt et al. 1975 | +0.05 | −7.1 | a |

[a] Insufficient data for faint state.

## 5. INTERPRETATION OF THE X-RAY AND OPTICAL DIAGRAMS

Hertz et al. (1992) and Dieters & van der Klis (2000) have discussed the fact that Sco X-1 does not move through its Z diagram at a uniform rate. The maximum rate at which the position in the Z diagram changes occurs at the top of the FB. As Sco X-1 moves down the FB, changes occur more slowly, reaching a minimum near the intersection of this NB and FB. As Sco X-1 then moves from this position onto the NB, its speed in the Z plot gradually increases and reaches a local maximum near the vertex of the NB and HB. From this point onto the HB, its motion appears to slowly decrease. More data are needed to confirm this latter trend.

The rate at which Sco X-1 moves through its Z diagram and the nearly linear relationship between $B$ and $S_Z$ deduced from the studies of Willis et al. (1980) and Vrtilek et al. (1991) provide a qualitative explanation for the morphology of the Sco X-1 $B$-magnitude histograms. When Sco X-1 is on its FB, an asymptotic behavior exists for the $B$ magnitudes with an approximate cutoff magnitude near 12.1. Movement along the FB is therefore expected to produce a distribution of $B$ magnitudes that is truncated near this value. Excursions up and down the FB, coupled with limited observing windows, account for the nonrepeatable value of the peak $B$ magnitude. Near the NB-FB vertex, Sco X-1 moves slowly. Observationally, this vertex occurs at a value of $B \sim 12.6$. Since Sco X-1 spends a relatively large amount of time near this location, a second peak in the $B$-magnitude histogram is created. On the NB, the $B$-magnitude distribution is truncated at fainter magnitudes by the rareness at which Sco X-1 is on its HB. The speed at which Sco X-1 moves along the NB is low and fairly constant. A broad $B$-magnitude distribution is therefore expected, and its peak would on average be located further away from the $B = 12.6$ peak than the peak associated with the FB. The observed NB peak is not expected to be identical from season to season, since it depends on the duration of the observing window and the detailed movements of Sco X-1 along its NB within that window.

As previously noted, Sco X-1 is rarely present on its HB. Based on the $S_Z$ versus $B$ relation, Sco X-1 is expected to be optically faint in this phase. The data presented by Mook et al. (1975) suggest that the HB $B$-magnitude distribution reaches a limiting value. The study by Hasinger et al. (1990) of the low-mass X-ray binary Cyg X-2 supports the view that when on the HB, X-ray and optical fluxes are directly related. These findings suggest that the faint Sco X-1 $B$ peak seen in the Mook et al. data and displayed in the 1970 panel in Figure 2 is quite likely associated with the HB. The slow movement of Sco X-1 during its HB phase, suggested by the work of Dieters & van der Klis (2000), would produce a fairly broad $B$-magnitude distribution, truncated at faint magnitudes by the asymptotic behavior mentioned above.

The Psaltis et al. (1995) model provides an explanation for the simultaneously observed Sco X-1 X-ray and optical behavior seen in Figure 3. During the low accretion rate phase, the electron temperatures of the magnetosphere and hot central corona rise as the kinetic energy of the infalling material is released. The system's luminosity increases, and both the X-ray and optical fluxes rise. This phase corresponds to the HB in the X-ray color-color plot. As the accretion rate continues to go up, the radiation pressure increases, causing the velocity of the infalling material to decrease and material to pile up near the neutron star. More X-ray photons are then absorbed and escape as optical photons. This phase corresponds to the NB, where the X-ray and optical magnitudes are inversely related. As the accretion rate increases beyond this value, the rate reaches and then exceeds the Eddington critical rate. The mass flow then becomes chaotic, with matter moving inward in some regions and outward in others. As material builds up in the outer corona, the electron scattering opacity increases and redirects some of the emitted X-ray photons onto the accretion disk. During this phase, which corresponds to the FB, the optical and X-ray fluxes are therefore correlated.

## 6. SUMMARY AND CONCLUSION

It has been known for many years that the relationship between the optical and X-ray emission from Sco X-1 is complex. Theories that its optical emission arises solely from X-rays reprocessed in an accretion disk have not been able to satisfactorily explain the observed relation between the optical and X-ray emission, or the multiple-peaked appearance of the $B$-magnitude histograms. The major findings of this paper are as follows:

1. The X-ray/optical behavior of Sco X-1 is adequately explained by the theoretical model of Psaltis et al. (1995). Patterns in the X-ray–$B$-magnitude plot correspond to different branches in the Z diagram. Self-absorption of X-rays by material close to the surface of the neutron star and the subsequent emission of this energy at lower energies accounts for the observed inverse relation between the X-ray and optical flux along the normal branch. The flaring branch is associated with super-Eddington mass flows where some of the escaping X-rays are reprocessed in the system's accretion disk.

2. The morphology of the $B$-magnitude histograms is related to movement in the Z diagram. The most frequently observed peak, at $B \sim 12.6$, is associated with the vertex of the normal and flaring branches. Sco X-1 moves slowly through this phase, thus accounting for the persistence of this feature in $B$-magnitude histograms. Peaks on either side of $B = 12.6$ are caused by differing velocities in the Z diagram, the temporal extent of the observational database, and the existence of an asymptotic $B$-magnitude behavior on the flaring branch and horizontal branch.

This research was supported by NSF grant HRD 96-28730. R. T. Z. gratefully acknowledges support from the New Mexico Alliance for Graduate Education and the Professionate through NSF grant HRD 00-86701. We also acknowledge valuable assistance from A. Harmon, the BATSE team, and W. Priedhorsky.